\documentstyle[psfig]{mn}
\newcommand{\kms}{\thinspace\hbox{$\hbox{km}\thinspace\hbox{s}^{-1}$}}
{\rm }\newcommand{\vsini}{\thinspace\hbox{$v\sin i$\ }}
\newcommand{\cai}{Ca\,{\sevensize I}}

\newcommand{\fei}{Fe\,{\sevensize I}}
\newcommand{\gamvels}{$\gamma$ velocities}
\newcommand{\gamvel}{$\gamma$ velocity}
\begin{document}
\title[$\gamma$ velocities of four dwarf~novae]{The systemic
velocities of four long-period cataclysmic variable stars}

\author[R.C. North, T.R. Marsh, U. Kolb, V.S. Dhillon, C.K.J. Moran]
{R. C. North$^{1,4}$, T. R. Marsh$^1$\thanks{email: trm@astro.soton.ac.uk}, U. Kolb$^2$, V. S. Dhillon$^3$ and C. K. J. Moran$^1$\\ 
$^1$University of Southampton, Department of Physics \& Astronomy, Highfield,
Southampton SO17 1BJ \\
$^2$The Open University, Department of Physics \& Astronomy, Walton
Hall, Milton Keynes, MK7 6AA \\ 
$^3$University of Sheffield, Department of Physics \& Astronomy,
Sheffield, S3 7RH\\
%$^3$University of Leicester, Astronomy Group, University Road,
%Leicester, LE1 7RH \\
%$^4$University of Sussex, Astronomy Centre, Brighton, BN1 9QJ \\
$^4$Met Office, London Road, Bracknell, Berkshire, RG12 2SZ
}
\date{Accepted 
      Received 
      in original form }
\pubyear{2001}

\maketitle

\begin{abstract}

Although a large number of orbital periods of cataclysmic variable
stars (CVs) have been measured, comparison of period and luminosity
distributions with evolutionary theory is affected by strong selection
effects. A test has been discovered which is independent of these
selection effects and is based upon the kinematics of CVs (Kolb
\& Stehle, 1996). If the standard models of evolution are correct then
long-period ($P_{\rm orb} > 5$hrs) CVs should be typically less than
1.5 Gyr old, and their line-of-sight velocity dispersion
($\sigma_\gamma$) should be small. We present results from a pilot
study which indicate that this postulate is indeed true. Four
long-period dwarf novae (EM~Cyg, V426~Oph, SS~Cyg and AH~Her) were
observed over a complete orbit, in order that accurate radial
velocities be obtained. We find values of -1.7, 5.4, 15.4 and 1.8
\kms\ with uncertainties of order $3$\,\kms, referred to the dynamical
Local Standard of Rest (LSR), leading to a dispersion of
$\sim$8\kms. Calculation of a 95 per cent confidence interval gives
the result $4 < \sigma_\gamma < 28$\kms\ compared to a prediction of
$15$\,\kms. We also have an improved determination of mass~donor
spectral type, $K_{2}$ and $q$ for the four systems.
\end{abstract}

\begin{keywords}
binaries: spectroscopic -- novae, cataclysmic variables -- accretion,
accretion discs -- techniques: radial velocities    
\end{keywords}

\section{Introduction}

Dwarf novae (DN) are a sub-group of cataclysmic variable stars. They
consist of a white dwarf (primary) star accreting matter from
a red dwarf (hereafter `mass~donor') which is in contact with its
Roche lobe. The material streams from the mass~donor through the inner
Lagrangian point into the potential well of the primary, where it
impacts onto the outer edge of an accretion~disc. Observable features
which distinguish DN from other CVs are the outbursts which occur on a
regular basis (with a baseline from days to months) and cause the
system to increase in brightness by $m_{v} \approx 2 - 5$ magnitudes.

%If conservative mass transfer from the less massive to the
%more-massive star is assumed to occur in CVs, then the separation,
%$a$, of the stellar components in the binary system should
%increase~\cite{patt,king}. In response to the mass loss the donor star will
%also re-adjust its radius, pulling it away from contact with the Roche
%lobe. At this point mass transfer should cease. 
The mass transfer process in CVs is driven by orbital angular momentum loss. 
%The solutions are to expand the star as it transfers mass, or to shrink
%the Roche lobe. The former is not supported by observations. The Roche
%lobe is most easily reduced in size by reducing the binary separation,
%which is most easily done by removing angular momentum. 
Gravitational radiation is highly effective at short orbital periods,
and theoretical calculations reproduce well the observed mass transfer
rates in CVs. For longer period systems, with higher mass transfer
rates, an alternative mechanism is needed. This is widely accepted to
be {\it magnetic braking}, a spin-down effect on the mass~donor due to
a stellar wind that effectively co-rotates with its magnetic
field. Due to effective tidal locking in CVs, the mass~donor will not
slow down, and instead angular momentum is extracted from the binary
orbit. However, magnetic braking still remains only a hypothesis, and
the level of direct observational support is still low.

%The present explanation for the existence of the so-called
%'period-gap' in the CV period distribution involves the concept of
%disrupted magnetic braking for those systems with a fully convective
%mass~donor. If the driving mechanism is interrupted the mass transfer
%time scale will exceed the time required for the mass~donor to adjust
%its radius to one appropriate to its lower mass. Since this radius is
%smaller than the Roche lobe radius, the star will detach and mass
%transfer ceases. The Roche lobe radius will decrease over time, as
%gravitational radiation or other angular momentum loss mechanisms will
%still operate. Therefore the Roche lobe will eventually shrink to the
%new size of the mass~donor, re-establishing contact and resuming mass
%transfer at some shorter orbital period.  This immediately introduces
%an age split into the CV population; those systems above the period
%gap must be younger than those below it. 
In the standard model for the formation and evolution of CVs magnetic
braking is only active for donor stars with a radiative core. Hence
there is a marked difference between the evolutionary time scale of
systems above the period~gap ($P_{\rm orb} \ga 3$\,h) and those below
the gap ($P_{\rm orb} \la 2$\,h). Kolb \& Stehle~\shortcite{kolste}
used population synthesis methods to confirm this. They determined the
age structure of a model population of Galactic CVs, and by convolving
this with the observed age-space velocity relation of
Wielen~et~al.~\shortcite{wie} (which is believed to result from the
diffusion of stellar orbits due to gravitational interactions with
massive objects) obtained the theoretical distribution of systemic,
hereafter $\gamma$, velocities. They showed that the age of a system
in the present CV population is largely determined by the time-scales
of the orbital angular-momentum loss mechanisms. In addition, the
models showed that the brightness-dependent selection effects which
have hitherto plagued comparisons between observations and theory
preserve the age differences, thereby providing an opportunity to test
directly the magnetic braking model.  So, if the standard models are
correct, then the CVs having periods longer than the upper limit of
the period~gap ($P_{\rm orb} \ge 3$\,hours) should be younger ($\le
1.5$Gyr), and therefore have a smaller line-of-sight velocity
dispersion according to the empirical age--velocity dispersion
relation (predicted value $\sim 15$\,\kms).

%due to them having experienced fewer gravitational interactions
Conversely, those CVs with orbital periods shorter
than the lower limit of the period~gap, should be older ($\ge$ 3-4Gyr)
and show a larger velocity dispersion (predicted as $\sim 30$\,\kms).
\begin{figure}
\hspace*{\fill}
\psfig{file=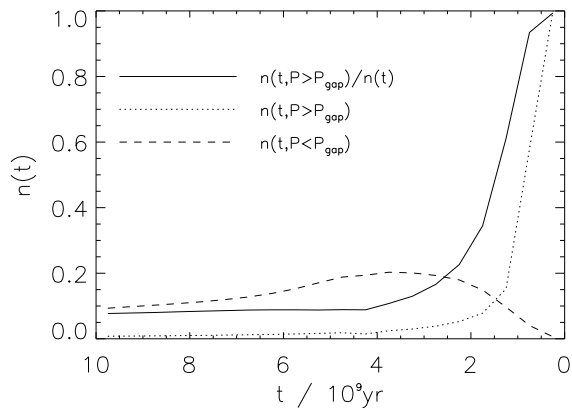,width=85mm}
\hspace*{\fill}
\caption{The figure (from Kolb \& Stehle, 1996) shows the relative
number of systems as a function of age, for CVs above the period gap
($P_{\rm orb}>3$hr, dotted line) and below the gap ($P_{\rm orb} <
2$ hr, dashed line). The solid line is the total fraction of CVs above
the period gap as a function of age. The relative youth of the longer
period systems is a firm prediction of the magnetic braking model of
CV evolution.}
\label{fig:ages}
\end{figure}

In a review by van Paradijs, Augusteijn \& Stehle~\shortcite{vanP},
the observed $\gamma$ velocities for a sample of CVs were collected
from published radial-velocity studies and statistically
analysed. They could detect no difference between the velocity
dispersions for below-gap and above-gap objects~(see
section~\ref{sec-dis}). All their data were taken from the literature,
and they assumed that there was no significant difference between
measurements taken from emission and absorption line measurements in
their sample.
%They concluded that for the non-magnetic CVs (of which they
%had a sample size of 72), the rms scatter of the velocity distribution
%is 33.4\kms. Assuming that the measurement error and the intrinsic
%dispersion add in quadrature, they calculated that the
%$\gamma$--velocities must have an intrinsic dispersion of $27$\kms,
%the measurement error having been deduced as $19.7$\kms (the rms
%scatter of a single $\gamma$ measurement). 
The problems associated with using values taken from the literature
are not negligible. The $\gamma$ velocity is a measure of the
centre-of-mass radial velocity of the binary star, and can be deduced
directly from radial velocity curves.  Typically with the emission
lines from CVs, a $\gamma$-value which is only one tenth the size of
the radial velocity variations in the system, and only one hundredth
of the total spectral line width, is being measured.  The radial
velocities obtained from emission lines are well known to be affected
by the internal motions in the disc or stream. Therefore, the
measurements of $\gamma$ obtained from emission lines may not
accurately reflect the motion of the white dwarf at all. This will be
especially true if the inner disc is distorted in some way, for
example, in magnetic CVs where the field on the primary is so strong
it disrupts the accretion disc and disrupts the accretion disc near to
the surface of the white dwarf. Measuring radial velocities from the
absorption lines due to the mass~donor helps reduce potential sources
of error. However, these absorption lines are not visible in all
dwarf~novae. They are generally only present in those systems with
orbital periods above the period~gap.
%Where two sets of spectral
%lines are present, radial velocity analyses can be carried out on both
%the absorption lines and the emission lines, thus enabling a
%calculation of the mass ratio, $q$, of the binary system.

Many radial-velocity curves for %both the white~dwarfs and mass~donors
%in many 
different CV systems have been constructed. However, if the only
requirement is a measurement of the radial velocity semi-amplitude,
care will not have been taken to ensure that systematic errors do not
dominate. It is vital that spectra are adequately sampled, and that a
reasonable number of radial velocity standards have been observed. If
a lower resolution is used to obtain data, then the probability of the
spectral lines being affected by blending is significantly increased.
%so using a dispersion which minimises this effect is
%important.
%Radial-velocity curves obtained from the
%absorption lines are inherently less prone to systematic error; those
%from the emission lines (to measure the motion of the white dwarf) are
%affected by the contribution from the accretion disc.
%So why can't we use existing archive data to complete this study? In
%short, the major problem is that measurement of the $\gamma$--velocity
%is usually {\it not} the primary aim of the study. This invariably
%means that care has not been taken to ensure that the spectra have
%been adequately sampled. Also, in most cases, not enough reliable
%radial-velocity standard stars will have been observed, introducing a
%further possible source of systematic error. These sources of error
%will plague any measurements of the systemic velocity. In several, a
%lower spectral resolution has been used, and so the probability of a
%spectral line being affected by blending is significantly increased.
%Unfortunately, obtaining accurate measurements of the $\gamma$
%velocities of CVs has always been inherently problematic. 
%We believe that these are not
%insurmountable problems for determining $\gamma$--velocities for dwarf
%novae. It is just that their measurement has not been the primary concern in
%previously published radial-velocity studies.

%%%%%%%

%%%%%%%

Here, we present the initial results of a project to deduce reliable
$\gamma$ velocities for a large sample of longer-period ($P_{\rm orb}
\ge 6$ hours) non-magnetic CVs. The aim of this initial investigation
was to test the methods and procedures described in \S III, to see
whether they could produce sufficiently accurate absolute
$\gamma$ velocity values, capable of being used to construct an
observed velocity-dispersion relation for Galactic CVs, and thus for
comparing directly with the theory.

%In \S II the observations and their reduction are briefly
%described. This is followed in \S III by presentation of the results,
%and their subsequent analysis. Finally, explanations for the observed
%results are discussed in \S IV, and the results of the paper are
%summarised in the conclusions, \S V.

%%%%%%%%%%%%%%%%%%%%%%%%%%%%%%%%%%%%%%%%%%%%%%%%%%%%%%%%%%%%%%%%%%%%%
\section{Observations}

During five nights starting from 19th June 1997, high time-resolution
spectroscopy was carried out using the Intermediate Dispersion
Spectrograph (IDS), on the 2.5-m Isaac Newton Telescope (INT) on the
island of La Palma, on four longer period ($P_{\rm orb} > 6$ hours)
DN. The objects observed are listed with exposure times and orbital periods
in Table~\ref{tab:info}. Conditions were clear throughout the five
nights of observing, and the seeing was approximately 1 arc
sec. Spectra of radial-velocity standard stars were also obtained to
calibrate the velocities measured from the DN. They were also used to
obtain a spectral type for the mass~donor in each DN, and so only
those with good spectral types were chosen. These were selected from
the lists of Marcy, Lindsay \& Wilson~\shortcite{mar}, Barnes, Moffett \&
Slovak~\shortcite{barnes}, Beavers~et~al.~\shortcite{beav}, Duquennoy, Mayor
\& Halbwachs~\shortcite{duq} and Eggen~\shortcite{egg}. The spectra covered the
wavelength range from $6220-6640$\AA, which includes absorption lines
from the mass donor around $6350-6500$\AA\ and the H$\alpha$ emission
line at 6563\AA. All the spectra had a dispersion of 0.4\,\AA\ per
pixel, resulting in a resolution with the 500-camera of 0.8\AA\
(FWHM). Each DN exposure was of the order of $\simeq$ 200 -- 300 seconds,
in order to keep sufficient phase resolution that no orbital smearing
would be introduced. The DN observed all had orbital periods in the
range $0\fd25 \leq P_{\rm orb}\leq 0\fd29$, and were tracked for a
minimum of 6 hours at one time. The 0.8 arc sec wide slit was rotated
in order to capture the spectrum of both the DN and a second star on
the slit, to correct for slit losses.

The CCD frames were bias-subtracted and flat-fielded. Variations in
the illumination of the slit were corrected using exposures of the
twilight sky. The object spectra were extracted using the
optimal weighting technique of Horne~\shortcite{horne}. Arc
calibration spectra were taken approximately every half hour, and with
each change of object. They were fitted with fourth-order polynomials,
which had an rms scatter of 0.005\AA. The spectra were corrected onto
a flux scale using a flux standard from Oke \& Gunn~\shortcite{oke}.
\begin{table}
\caption{A summary of the observations taken in June 1997 on the 2.5-m
INT. All used the IDS 500-mm camera with the Tek CCD chip and R1200Y
grating giving a dispersion of 0.4\AA\ per pixel, and resulting in a
resolution of 0.8\AA\ (FWHM).}
\label{tab:info}
%\begin{minipage}{140mm} 
\begin{center}
\begin{tabular}{lccc}
\hline
Object   &  Date observed & $T_{exp}$  & $P_{\rm orb}$$^{\rm Reference}$ \\
         &(UT, June 1997)&  (s)  &  (days) \\
\hline
EM Cyg   &  21.93-22.22  & 200 & 0.290909$^1$  \\
V426 Oph &  20.92-21.21  & 200 & 0.285314$^2$  \\
SS Cyg   &  23.01-23.23  & 200 & 0.27512973$^3$  \\
AH Her   &  19.91-20.17  & 300 & 0.258116$^4$  \\
\hline
\hline
\end{tabular}

\smallskip
References: (1) Stover et al., 1981 ; (2) Hessman, 1988; \\ 
(3) Hessman et al., 1984; (4) Horne, Wade \& Szkody, 1986.    
\end{center}
%\end{minipage}
\end{table}

%%%%%%%%%%%%%%%%%%%%%%%%%%%%%%%%%%%%%%%%%%%%%%%%%%%%%%%%%%%%%%%
\section{Methods}
\label{sec-methods}

%The data for each of the DN was dealt with using the same method. 
%To obtain a value for the velocity dispersion of the order 15\kms, it
The $\gamma$ velocities needed to be measured to within 5\,\kms\ so
that a velocity dispersion of the order 15\kms\ could be detected. The
following process illustrates how this was achieved, and how potential
sources of systematic error
%(which are the main problem using a procedure like this) 
were minimised.
\begin{enumerate}
        \item First, the individual spectra of each DN were re-binned
        onto a logarithmic wavelength scale~\cite{stov80}.  Then each
        individual DN spectrum was cross-correlated with that of a
        radial-velocity standard (re-binned onto an identical scale),
        following the method of Tonry \& Davis~\shortcite{td}, in
        order to determine heliocentric radial velocities.
        Adjustments were made at this stage to allow for the
        radial-velocity of the standard star. No allowance (at this stage) was
        made for the rotational broadening of the absorption lines
        from the mass~donor. 
        \item The radial velocities were then fitted with a circular
        orbit fit of the form 
\begin{equation}
 V = \gamma + K\sin\frac{2\pi(t - t_{0})}{ P_{\rm orb}} 
\end{equation} 
        to calculate the radial-velocity semi-amplitude (K), the
        systemic velocity ($\gamma$), and the phase-zero point where
        phase zero, $\phi_{0}$, is defined as the phase at which the
        radial velocity of the mass~donor crosses zero moving from
        negative to positive~(see Section~\ref{sec-rvs}). These
        initial radial velocity fits were made to allow the orbital
        motion of the mass~donor to be removed from the individual
        spectra in order to create an average spectrum from which to
        determine the spectral type.  
        \item Repeated (ii) using radial-velocity standards
        artificially broadened to match the widths of the absorption
        lines seen in the DN. Measurements of the rotational
        broadening of the absorption lines are discussed in
        section~\ref{sec-vsini}. %Determination of the broadening also
%       gave an estimate of the spectral type of the mass~donor.

        \item An estimate of the mass donor spectral type could be
        deduced simply by inspection of the spectra~(see North et al., 
        2000, and section~\ref{sec-sptypesec})
%       The behaviour of several of the absorption lines is
%       obviously closely related to temperature in the standard stars
%       observed. Casares, Charles, Naylor \&
%       Pavlenko~\shortcite{cascha} noted the usefulness of several
%       relative line depths in constraining the spectral type of the
%       mass donor in V404~Cyg over a similar wavelength region to the
%       one being used here.
%       The spectral-type of the mass~donors could also be deduced in
%       the following way.  A range of standard stars with different
%       spectral types were artificially broadened, and then fractions
%       of those standard spectra were subtracted from the average DN
%       spectrum (with the orbital motion of the mass~donor
%       removed). The fraction of a particular standard which produced
%       the lowest $\chi^{2}$ value determined the best-fit spectral
%       type.
        An alternative method used is discussed in
        section~\ref{sec-sptypesec} together with the results
        obtained.
        \item Finally, the cross-correlation was repeated again, using
        the radial-velocity standard (of best-fitting spectral-type)
        broadened to the measured rotational velocity of the mass
        donor.
\end{enumerate}
To avoid introducing errors due to spectral-type mismatch between
standard star and DN, steps (i) to (v) were repeated with velocity
standards of several different spectral types (ranging from
G8V--M6V). The radial velocities obtained for each DN did not differ
significantly, implying that this is not a large source of systematic
error here. %This process also identified those radial-velocity
%standards which are good to use for this purpose, and those which may
%still have an uncertain radial-velocity value or spectral type.

%%%%%%%%%%%%%%%%%%%%%%%%%%%%%%%%%%%%%%%%%%%%%%%%%%%%%%%%%%%%%%%%%%%%%%
\section{Results}

In this section, the results obtained using the methods described in
section~\ref{sec-methods} are presented. Section~\ref{sec-rvs}
describes the radial velocities obtained for each mass~donor. This is
followed by the results of the rotational broadening measurements
(section~\ref{sec-vsini}), and the mass-donor spectral typing process
(section~\ref{sec-sptypesec}). The resulting $\gamma$ velocities are
given in section~\ref{sec-gamma}. Finally Doppler maps of the
H$\alpha$ emission are presented in section~\ref{sec-dopp}.

%%%%%%%%%%%%%%%%%%%%%%%%%%%%%%%%%
\subsection{Mass~donor radial velocity curves}
\label{sec-rvs}

\begin{figure*}
\hspace*{\fill}
\psfig{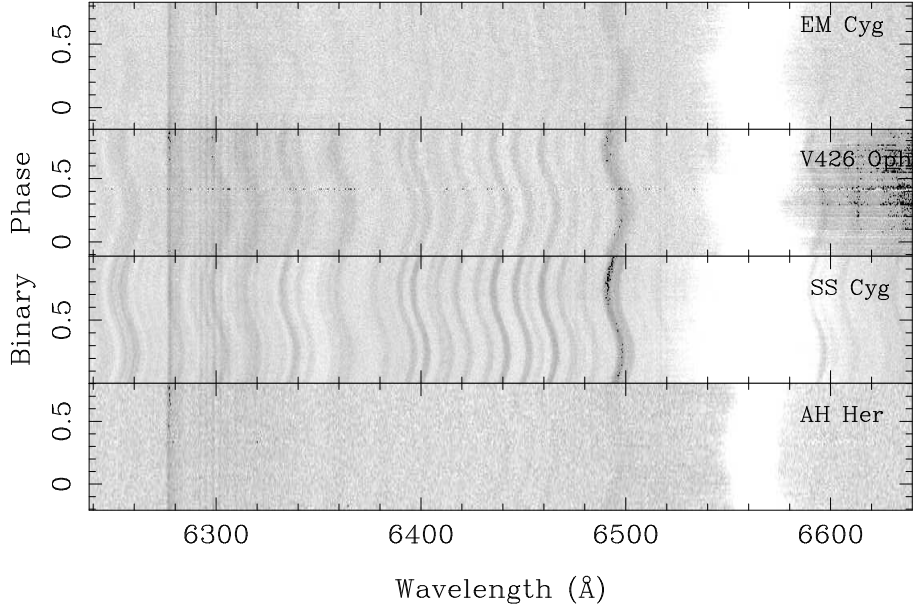}
\hspace*{\fill}
\caption{Trailed spectra (wavelength versus binary phase) of the DN
observed, scaled to $\pm0.2$ of the continuum level. The
H\protect{$\alpha$} lines have been saturated (white) in order to
bring up the contrast of the absorption lines from the mass donor,
which can be seen in the wavelength range 6300--6500\,\AA. From
top-to-bottom, the panels show the systems in descending orbital
period length.  In V426~Oph and SS~Cyg, the sinusoidal nature of the
absorption lines (black) is clearly visible. The lowest panel contains
the trailed spectra of AH~Her. The mass donor lines are not as clear
as in all the other dwarf novae, but the feature at \protect{$\lambda$
6495\,\AA} is visible. The absorption lines in EM~Cyg have been
corrected for the presence of an additional late-type spectrum (see
North~et~al.,~2000). See \protect{Fig.~\ref{fig:dopp}} for
H\protect{$\alpha$} detail.}
\label{fig:trails}
\end{figure*}

\begin{figure*}
\hspace*{\fill}
\psfig{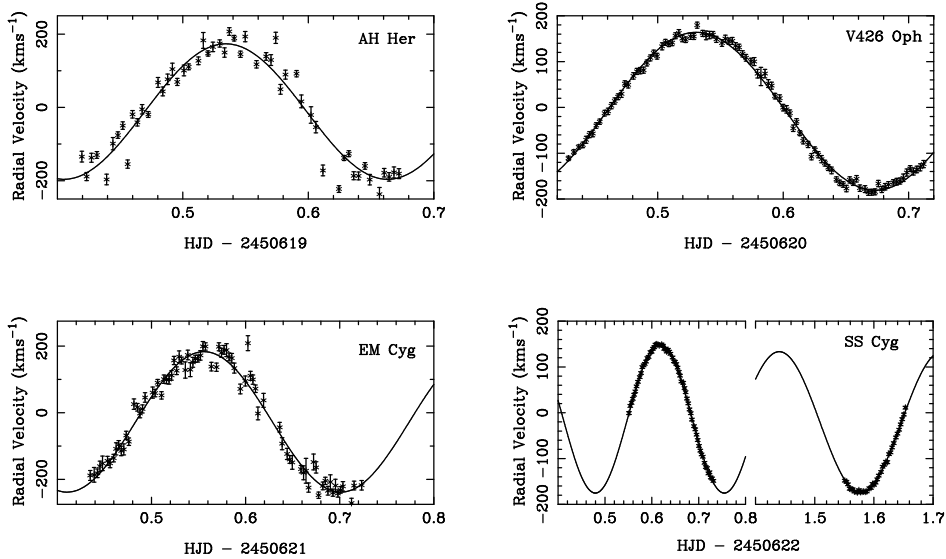}
\hspace*{\fill}
\caption{Sinusoidal fits to the radial velocity data for each dwarf
nova. Clockwise from top left, AH~Her, V426~Oph, EM~Cyg and
SS~Cyg. The data from two nights worth of observations for SS~Cyg is
combined into the lower right-hand plot.}
\label{fig:curves}
\end{figure*}

To obtain radial velocities, the absorption spectrum from the
mass~donor, which can be seen in the wavelength region
$\lambda\lambda\,6350-6540\,$\AA\ was used. The absorption lines from
the mass~donor are easily distinguished from interstellar features, as
they exhibit a sinusoidal variation. The trailed spectra in
Fig.~\ref{fig:trails} show the behaviour of the spectral lines over a
full orbital period as observed in each system.
Figure~\ref{fig:trails} clearly shows the sinusoidal behaviour of the
absorption lines. Each panel also shows how the emission lines vary
approximately 180\degr\ out of phase with the absorption lines. The
lines at $\approx 6300$\,\AA\ are interstellar absorption. The
difference between the trails of the individual systems are apparent
in this plot. For example, the absorption lines in AH~Her are barely
visible, whereas those in SS~Cyg are very obvious. 
%DN with orbital periods greater than $\approx 5$ hours have
%mass~donors with spectral types in the early-to-mid K range. Due to
%their higher temperatures they can appear in the composite spectrum at
%shorter wavelengths than their M-dwarf counterparts, which tend to get
%quenched by the accretion~disc spectrum in the visible wavelength
%region. The relative contribution to the continuum by the
%accretion~disc at the time of the observations also affects the
%visibility of the absorption lines. If the disc is bright (in
%outburst) then the absorption lines may not be visible. 
All the objects were observed when the system was in or approaching a
quiescent disc state. Also, the strength of the absorption lines may
vary around the binary orbit if the hemisphere of the mass-donor
facing the white dwarf is being irradiated by the accretion~disc and
boundary layer. This manifests itself as a departure from a
circular orbit fit to the radial velocities around phase~0.5 (see
Fig.~\ref{fig:curves}). The behaviour of the H$\alpha$ line is also
seen to differ markedly in each object. This is discussed in
section~\ref{sec-dopp}.

%The region of spectrum used for the cross-correlation was between
%$6350 - 6540$\AA. This wavelength range contains absorption lines
%from low-ionisation states of calcium and iron, which originate at the
%mass~donor. The region containing interstellar features was avoided,
%as was that close to the H$\alpha$ line wings. 
%Separate radial velocity curves were obtained for each of the
%radial-velocity standards used (spectral types ranging from G8V--M6V),
%to check that spectral-type mismatch did not produce a significant
%error.  
The radial velocities resulting from the cross-correlation procedure
were fit with a sine curve, to determine the semi-amplitude, $K_2$,
phase-zero point, $\phi_{0}$, and $\gamma$ velocity.
\begin{table}
\caption{Orbital parameters determined from the absorption line radial 
velocity curves}
\label{tab:params} 
\begin{center}
%\hspace*{\fill}
\begin{tabular}{lcc}
\hline
Object & $K_2$  &  $\phi_{0}$$^{a}$   \\
       & (\kms) &  2450000+     \\
\hline\hline
EM Cyg & 202$\pm$3 & 621.4833(5)   \\
V426 Oph & 179$\pm$2 & 620.4596(3)  \\
SS Cyg & 165$\pm$1 & 622.5483(2)    \\
AH Her & 175$\pm$2& 619.4669(5)     \\
\hline
\end{tabular}

\smallskip
$^{a}$\, Figure in brackets is the error on the final digit.
%\vspace{-7mm}
\end{center}
%\hspace*{\fill}
\end{table}
The orbital~periods are already well-determined for the four observed
DN, so we use the values given in the literature (see
Table~\ref{tab:info} for details). 
%Then the best-fit spectral type of
%the mass~donor in each DN was determined (as in
%section~\ref{sec-sptypesec}), as was the level of rotational
%broadening of the absorption lines (\vsini, section~\ref{sec-vsini}).
%Finally, the radial velocities were re-calculated using the standard
%with best-fit spectral type, whose absorption lines had been
%artificially broadened to the best-fit value of \vsini.
Figure~\ref{fig:curves} shows the sinusoidal fits to the radial
velocities of each DN. The radial-velocity standard used to obtain
each fit was of best-fitting spectral type and artificially broadened to
the value obtained in section~\ref{sec-vsini}.

%%%%%%%%%%%%%%%%%%%%%

\subsection{Projected rotational velocities of the mass~donors} 
\label{sec-vsini}

\begin{figure}
\hspace*{\fill}
\psfig{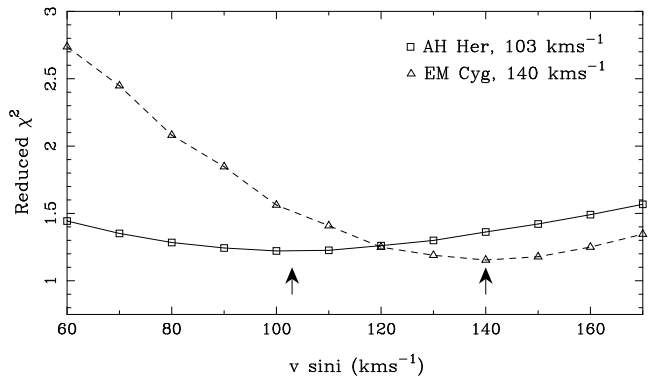}
\hspace*{\fill}
\caption{Examples of the \protect$\chi^{2}$ versus \protect\vsini\ curves
obtained during the rotational broadening measurement procedure. The
arrows denote the positions of the minima of the curves (optimum
\vsini\ value).}  
\label{fig:vsini}
\end{figure}

The rotational broadening of the absorption lines was determined by
artificially broadening the radial velocity standards to values
between 0 and 200\,\kms\ (using increments of 10\kms). The orbital
motion of the mass~donor was removed from the individual CV spectra,
and the results rebinned onto a uniform velocity scale. The resulting
spectra were then co-added. The fraction of each broadened standard
which best removed the mass-donor lines was then calculated, using an
optimisation technique which minimised the scatter between the
spectrum of the standard and the DN spectrum. Plotting $\chi^{2}$
versus the amount of artificial broadening used on the standard
produces a minimum in $\chi^2$ at the optimum value of
\vsini\ (see Fig.~\ref{fig:vsini} for examples). This gave a value
for the rotational broadening of the mass~donor.  
In order to check that choosing one spectral type for the mass~donor
did not introduce a further source of systematic error, we repeated
the \vsini\ measuring technique with all the available standard
stars. The standard error on the mean of the sample was $2\kms$,
implying that the choice of standard used to calculate \vsini\ was not
significant. A value of 0.5 was used for the limb-darkening
coefficient for each DN. Changing this value over the range $0-1$ did
not alter the deduced values of \vsini\ by more than 5\kms, see
North~et~al.~\shortcite{north} for details of the method
followed. Table~\ref{tab:vsini} shows the measurements of \vsini\ for
each of the observed systems.  Horne,~Wade~\&~Szkody~\shortcite{hws}
calculated a value for the rotational broadening of the mass~donor in
AH~Her of 112$\pm$17\kms. Our value of 103$\pm$2\kms\ agrees with
theirs. Hessman~\shortcite{hessman} determines a value for the
mass~donor of SS~Cyg of 87$\pm$4\kms. Our value of 96$\pm$3\kms\
agrees at the 2$\sigma$ level.

\begin{table}
\caption{Results of the procedure to find the value of the rotational
broadening of the mass~donor in the four DN.}
\label{tab:vsini} 
\hspace*{\fill}
\begin{tabular}{lccc}
\hline
Object & \vsini\   & Spectral type & mass ratio\\
       &  (\kms)   & of standard & from \vsini \\
       &           &  used        & and K$_{2}$ \\
\hline
\hline
EM~Cyg   & $140\pm2$ &  K3V & $0.88\pm0.03$\\
V426~Oph & $132\pm3$ &  K5V & $0.96\pm0.01$\\
SS~Cyg   & $96\pm3$  &  K5V & $0.68\pm0.02$\\
AH~Her   & $103\pm2$ &  K7V & $0.69\pm0.02$\\
\hline
\end{tabular}
\hspace*{\fill}
\end{table}
%%%%%%%%%%%%%%%%%%%%%%%%%%%%%%%%%%%%%%%%%%%%%%%

\subsection{Mass donor spectral types}
\label{sec-sptypesec}

The radial-velocity standard spectra were each artificially broadened
to an appropriate value (that determined in section~\ref{sec-vsini}).
Then the fraction of that broadened standard which, when subtracted
from the DN spectrum best removed the spectral lines of the mass~donor
was calculated. The routine used minimised $\chi^{2}$ between the
DN spectrum (with a specific fraction of the chosen standard
subtracted) and a smoothed version of itself.  The standard producing
the lowest value of $\chi^{2}$ gave an indication of the mass~donor
spectral type. All the objects observed have mass~donors with a
spectral type later than K3. AH~Her best fitted a K7V
mass~donor. SS~Cyg and V426~Oph were both best-fit using a K5V
spectral type. Finally, EM~Cyg resulted in a K$3-5$ type mass donor
(see North~et~al.~\shortcite{north}).
\begin{table*}
\caption{The mass~donor spectral type determined for each
object. Previous estimates of the spectral type (from the
literature) are also given.}
\label{tab:sec} 
\begin{center}
%\hspace*{\fill}
\begin{tabular}{lcccc}
\hline
Object & Spectral Type &  Previous   & Fraction contribution& reference for \\
       & of mass~donor &  Spectral type & of mass~donor & previous type$^{a}$  \\
\hline\hline
EM Cyg & K3    &  K5V    &0.231$\pm$0.005&  3 \\
V426 Oph & K5V &  K3V    &0.336$\pm$0.006&  2 \\
SS Cyg & K5V  &  K2-K3V  &0.685$\pm$0.004&  4 \\
AH Her  & K7V  &  K0-K5V &0.102$\pm$0.006&  1 \\
\hline
\end{tabular}

\smallskip
$^{a}$\, 1) Horne, Wade \& Szkody,~1986; 2) Hessman,~1988; 3)
Stover~et~al.,~1981; 4) Martinez-Pais~et~al.,~1994.
%\vspace{-7mm}
\end{center}
%\hspace*{\fill}
\end{table*}

%An estimate of the spectral type of the mass~donor in each DN was
%carried out. A reliable spectral-type value was obtained simply by
%inspection of the spectra in the wavelength region $6350-6540$\,\AA,
%as noted in North~et~al.~\shortcite{north}. Casares, Charles, Naylor
%\& Pavlenko~\shortcite{cascha} determined useful spectral typing
%constraints for this wavelength region which were used as a guide.
%The depth of the \fei\ blend at $\lambda\lambda\/6400.0 + 6400.3$,
%compared to that of the \cai\ line at $\lambda\lambda\/6439.1$ varies
%rapidly with effective temperature over the spectral-type range in use
%(G8V--M6V). Casares~et~al.~\shortcite{cascha} noted some useful
%spectral-typing line relationships using a similar wavelength
%region. Specifically, they note that the relative depths between \cai\
%$\lambda\/6439.1$ and the blends \cai\ $\lambda\/6462.6 + {\rm
%Fe}\,${\sevensize I}\ $\lambda\/6462.7$ are particularly useful. They
%measured the equivalent widths of these lines, plus those of
%Fe\,{\sevensize I}\ $\lambda\lambda\/6430.9$, Fe\,{\sevensize I}\
%$\lambda\lambda\/6392.5+6393.6$ and Fe\,{\sevensize I}\
%$\lambda\lambda\/6400.0 + 6400.3$, and remarked that the ratios
%$\lambda\lambda\/6439/6431$, and the $\lambda\lambda\/6450/6439$ are
%very sensitive to $T_{\rm eff}$ over the spectral-type range in
%question.
\begin{figure*}
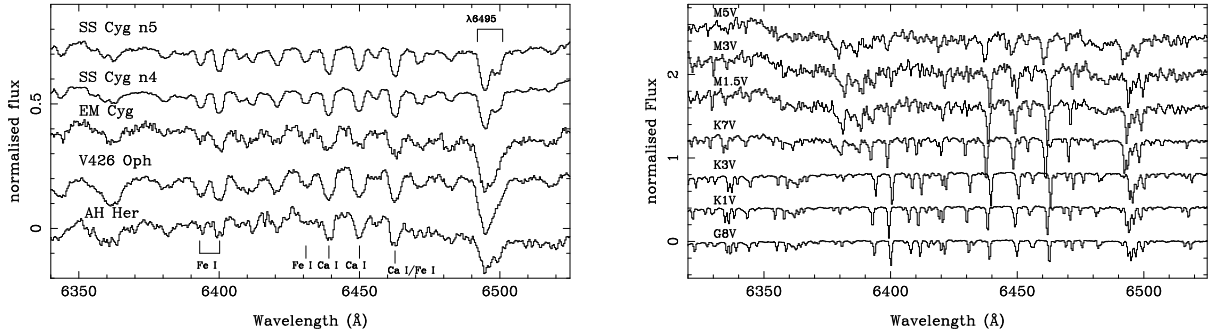

\hspace*{\fill}
\psfig{file=figure5.ps,width=75mm}
\hspace*{\fill}
%\caption{The average spectra of the mass~donors
%of the four DN observed. The raw spectra for each DN have had the
%orbital motion of the mass~donor removed, then been re-binned onto a
%uniform velocity scale and averaged}
%\label{fig:sec}
%\hspace*{\fill}
\psfig{file=figure6.ps,width=75mm}
\hspace*{\fill}
%\caption{A selection of the radial velocity standards observed,
%ranging from the earliest spectral type observed (G8V) to the latest
%(M5V).}
\caption{{\bf Left.} a) The average spectra of the mass~donors
of the four DN observed. The raw spectra for each DN have had the
orbital motion of the mass~donor removed, then been re-binned onto a
uniform velocity scale and averaged. {\bf Right} b) A selection of the
radial velocity standards observed, ranging from the earliest spectral
type observed (G8V) to the latest (M5V)}
\label{fig:temps}
\end{figure*}
Figure~\ref{fig:temps}a shows the average spectra of the four DN,
created by removing the orbital motion of the mass~donor from each
individual spectrum, re-binning onto a uniform velocity scale and
co-adding the results. The absorption lines are obviously broadened
(c.f. those of Figure~\ref{fig:temps}b); measured in each case to be
$\approx 100\kms$, see Table~\ref{tab:vsini}. Marked on the plot are a
few of the spectral lines found most useful in the spectral typing
process. The blend at $\lambda\/6495$ is also marked on the plot. It
is a blend of mostly \cai\ $\lambda\/6493.8$ and
\fei\ $\lambda\lambda\/6495.0,6495.7$ and $\lambda\/6496.5$. However,
it is the one absorption feature originating from the mass~donor which
is clearly visible in the trailed spectra of each object.  Purely by
inspection, there appear to be four reliable indicators of spectral
type in this wavelength range (see Fig.~\ref{fig:temps}a). First, there
appears to be a relationship between the relative depths of
$\lambda\/6439.1$ and the blend at $\approx\lambda\/6400$ which
changes rapidly over the spectral type range we are interested in
here. 
%This enables us to identify whether the spectral type is between
%G8 and K3, or between K5 and M6. 
Secondly, the \fei\ lines around $\lambda\/6400$ begin to change very
obviously around spectral type K7. The general shape of the spectrum
around 6370\AA\ allows us to deduce whether the mass~donor is a K- or
M-type star. Finally, the two spectral features just short of the
$\lambda\/6495$ blend vary significantly relative to each other over
the spectral range G8--M6. Figure~\ref{fig:temps}b shows a range of
typical standard star spectra covering spectral types G8V to M5V for
comparison.

%%%%%%%%%%%%%%%%%%%%%%%%%%%
\subsection{Deducing the $\gamma$ velocities}
\label{sec-gamma}

There have been many radial-velocity studies on DN which have
published $\gamma$ velocity values, both for DN in outburst and
quiescence. Several of these results do not appear consistent with
each other. We believe that this is just a symptom of the data
acquisition methods. Obtaining $\gamma$ velocities is not the primary
aim for many of these studies, and so it is highly probable that the
spectra are undersampled and therefore unsuitable for deducing
absolute \gamvel\ values. 
%We compiled the published $\gamma$ velocity
%measurements for the DN discussed in this paper. 
%SS~Cyg is a prime
%example of a single object with discrepant values for the
%\gamvel. Echevarria~et~al.~\shortcite{eche} obtain $-15.7\pm0.3$\kms
%for $\gamma$, from the absorption lines whilst SS~Cyg was at the end
%of an outburst (using echelle
%spectra). Friend,~Martin,~Smith~\&~Jones~\shortcite{fmsj} used the
%Na\,{\sevensize I} lines around 8190\AA\ to calculate a \gamvel\ of
%$-14\pm3$\kms. At the time of their observations, SS~Cyg was in
%5quiescence almost mid-way between two
%outbursts. Kiplinger~\shortcite{kip}, calculated a
%\gamvel\ of $-3\pm4$\kms, again from the absorption lines. Our
%heliocentric $\gamma$ value for SS~Cyg is $-14\pm2$\kms. If this is
%the case for SS~Cyg, one of the brightest DN for which it is easy to
%obtain measurements, then it can only be worse for the rest.

\begin{figure*}
\hspace*{\fill}
\psfig{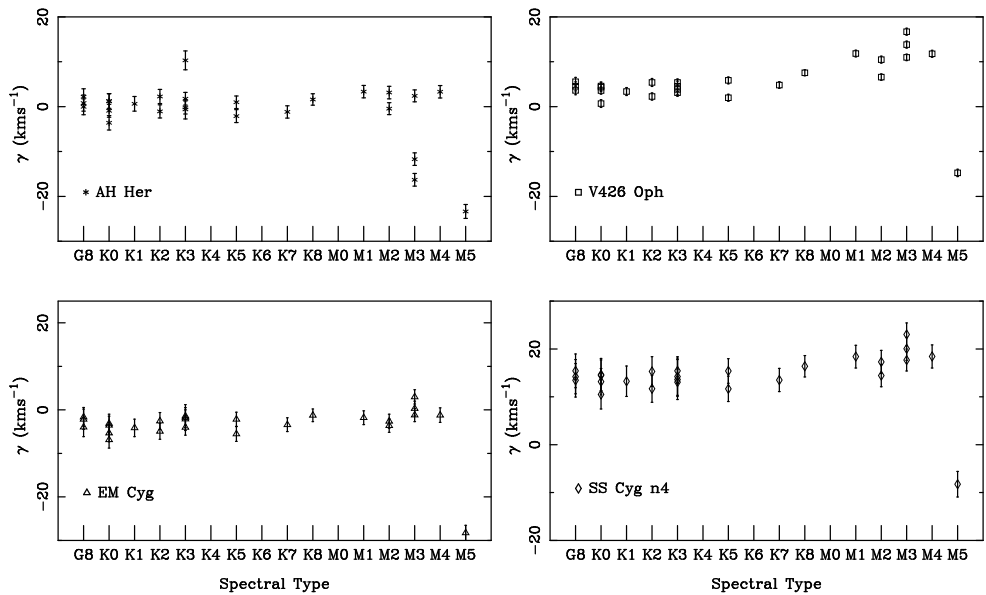}
\hspace*{\fill}
\caption{The \protect$\gamma$ velocities calculated with each
spectral type standard.}
\label{fig:gammas}
\end{figure*}

%
%We are using the absorption lines from the mass~donor for the radial
%velocity analysis and so observations were obtained when the
%contribution from the accretion disc was at a minimum (quiescence), in
%order to reduce systematic effects. Spectroscopically, this state can
%be identified in DN when the lower members of the Balmer series of
%spectral lines are in emission. All four DN were observed with strong
%H$\alpha$ emission.
\begin{table}
\caption{Heliocentric $\gamma$ velocities for four dwarf novae, and
corrections for the solar motion}
\label{tab:rvs} 
\hspace*{\fill}
\begin{tabular}{lccc}
\hline
Object & Heliocentric & Correction & Absolute\\ & $\gamma$ velocity &
for solar & $\gamma$ velocity \\ & (\kms) & motion (\kms) & (\kms) \\
\hline
\hline
EM~Cyg   & -27.38$\pm$1.88 & -25.65 & -1.74$\pm$1.9 \\
V426~Oph & -11.07$\pm$0.84 & -16.46 & 5.4$\pm$0.8 \\
SS~Cyg   & -13.09$\pm$2.88 &  -28.52 & 15.4$\pm$3.0 \\
AH~Her   & -5.99$\pm$1.54  & -7.75  & 1.8$\pm$1.5 \\
\hline
\end{tabular}
\hspace*{\fill}
\end{table}
The $\gamma$ velocities are calculated at the point in the analysis
described previously when the final radial-velocity curves are derived
for each object. They are deduced using a sinusoidal fit to the radial
velocities around the complete orbit, which have themselves been
determined using a standard of the same spectral type as the
mass~donor, which has been artificially broadened to the measured
\vsini. The values calculated at this point are heliocentric, and in
order to deduce the velocity dispersion for these measurements, it was
necessary to correct for the solar motion. For this we used the
dynamical Local Standard of Rest (LSR) as the reference
point~\cite{rv}. Figure~\ref{fig:gammas} shows the range of $\gamma$ values
obtained for each spectral type of radial-velocity standard, for each
DN. It was here that using so many radial-velocity standard stars was
most useful. Table~\ref{tab:rvs} shows the best-fitting heliocentric
$\gamma$ velocity values for each object, and the velocity shift 
applied (in \kms) to correct for the solar motion.
Most of the derived $\gamma$ velocity values for each dwarf~nova are
consistent to within the error margin of $\pm$2\kms. 
%The ones that do not agree to within this precision were found to be
%bad identifications in the observing log, and so the radial-velocities
%calculated with that standard had been corrected to the wrong
%zero-point.

Considering only the data obtained here, the velocity dispersion,
$\sigma_\gamma$, appears to be approximately 8\kms, smaller than the
value of 15\kms\ predicted by Kolb \& Stehle for longer period
systems, yet still consistent with it. We calculate a 95 per cent
confidence limit on the value of $\sigma_\gamma$ to be $4 <
\sigma_\gamma < 28$\,\kms, a large range with only 4 systems. If the
dispersion we measure is assumed to remain at the same magnitude but
the number of measured systems increases to 8 then the 95 per cent
confidence limit decreases to $5 < \sigma_\gamma < 16$\,\kms.

In addition system parameters for the four DN were calculated using
the derived values of \vsini\ and $K_2$ (see Table~\ref{tab:syspar}). 
\begin{table}
\caption{System parameters determined using the derived values of
\protect{\vsini\ } and $K_2$}
\label{tab:syspar} 
\begin{center}
\begin{tabular}{lccc}
\hline
Object & $M_2\sin^{3}i$  & $M_1\sin^{3}i$ & Inclination\\ 
& ($M_{\odot}$) & ($M_{\odot}$) & ($\degr$)\\
\hline
\hline
EM~Cyg   & 0.77$\pm$0.08  & 0.88$\pm$0.05 & 67$\pm$2$^1$ \\
V426~Oph & 0.62$\pm$0.02  & 0.65$\pm$0.02 & 59$\pm$6$^2$ \\
SS~Cyg   & 0.25$\pm$0.01  & 0.36$\pm$0.01 & 37$\pm$5$^2$ \\
AH~Her   & 0.28$\pm$0.01  & 0.41$\pm$0.01 & 46$\pm$3$^2$ \\
\hline
\end{tabular}

\smallskip$^1$ North~et~al.~(2000), $^2$~Ritter~\&~Kolb~(1998)
\end{center}
\end{table}

%%%%%%%%%%%%
\subsection{A note on the space velocity of SS~Cyg}
\label{sec-par}

One of the observed DN (SS~Cygni) has had its parallax measured using the Fine
Guidance Sensor (FGS) on the Hubble Space Telescope (HST)~\cite{harri1}. The
parallax measurement of $6.02\pm0.46\,$mas lead to a new distance for SS~Cyg of
$166\pm12\,$pc, almost twice the distance obtained previously. The
tangential velocity calculated using this distance is $93\,$\kms. So,
although our radial velocity agrees with theoretical expectations, the
total space velocity of $94\,$\kms\ does not. 

It is possible that there is a problem with the SS~Cyg parallax -- for
example Schreiber \& G\"{a}ensicke have shown that the small parallax,
and consequent implied luminosity causes problems with the disc
instability theory. It causes us problems here too, although even if
the distance were only $100\,pc$ the space velocity would still be a
relatively high $56\kms$. As SS~Cyg is only one system, little can be
deduced; it is clearly of interest for more parallaxes and proper
motions to be measured.

%%%%%%%%%%%%%%%%%%%%%%%%%%%%%%%%%%%%%%%%%%%%%%%%%%
\subsection{Doppler Tomography}
\label{sec-dopp}

\begin{figure*}
\hspace*{\fill}
\psfig{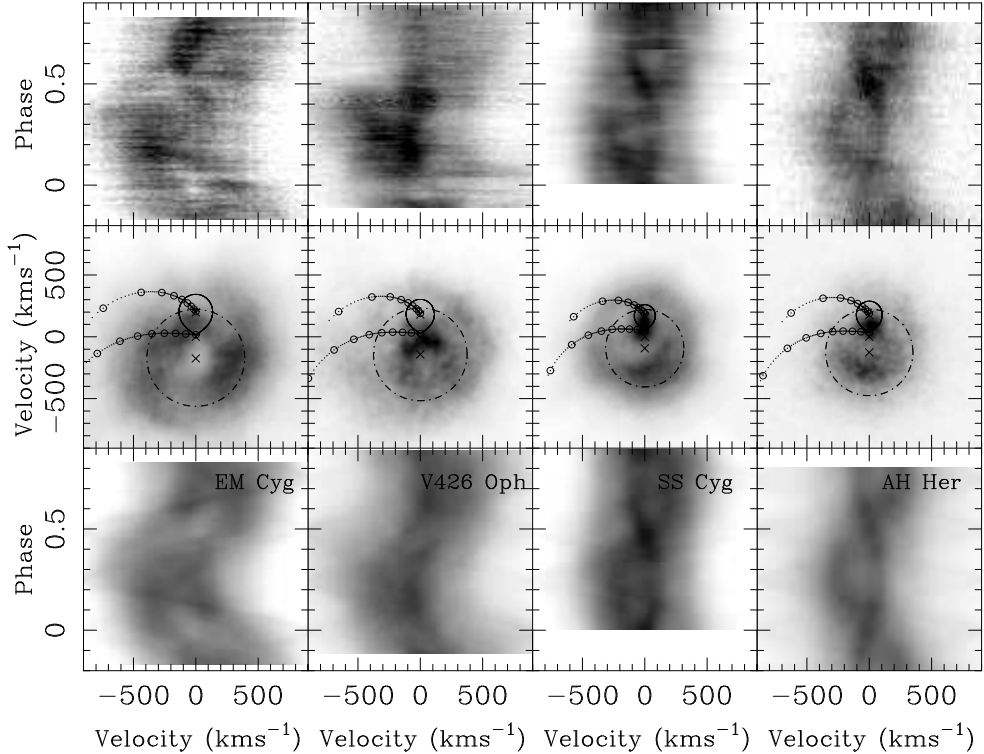}
\hspace*{\fill}
\caption{Doppler maps of the four dwarf novae. From left to right:
EM~Cyg, V426~Oph, SS~Cyg and AH~Her. From top to bottom, the panels
show a) the actual data (in the form of trailed spectra); b) The
Doppler map of the \protect{H$\alpha$} emission; c) The data
reconstructed from the Doppler map. The Roche lobes were plotted onto
each Doppler map using the mass ratios given in
\protect{Table~\ref{tab:vsini}} and the values of $K_2$ calculated in
section~\protect{\ref{sec-rvs}}. The predicted path of the gas stream
is also shown, and open circles are marked on at $0.1$\,$R_{L_1}$
intervals. The upper path is the Keplerian velocity along the stream;
the lower path is the direct velocity along it}
\label{fig:dopp}
\end{figure*}

A useful method of imaging the accretion discs in CVs is Doppler
tomography~\cite{mh}. This procedure maps emission from the CV in
velocity space, using a coordinate system which co-rotates with the
binary star. Due to the synchronous rotation of the mass~donor with
the binary motion, the shape of the Roche lobe is conserved in
velocity space, and can be plotted on the Doppler map.  The aim of the
observations, to measure accurate absolute \gamvels, means that these
data are ideal for applying Doppler tomography methods, in this case
to the H$\alpha$ emission line. Figure~\ref{fig:dopp} is a compilation
of the Doppler maps of the H$\alpha$ emission for the four
systems. The uppermost panel shows the observed trailed spectra
centred at $6562.760$\AA, using a velocity scale on the $x$-axis with
a range of $\pm 900\kms$. The middle panel in each column plots the
Doppler image. Finally, the lower panel presents the computed trailed
spectra, as reconstructed from the Doppler images. The observed data
are generally reproduced well, however several asymmetric features
exist in the observed data which the software tried to fit and failed.

Starting with the plots for AH~Her (column~four of
Fig.~\ref{fig:dopp}), it can be seen in the trailed spectra that there
are two obvious `S-wave' components. One of these is from the
mass~donor (starting at $-100$\,\kms\ at phase~$-0.2$). It manifests
itself in the Doppler map at velocities indicating an origin on the
irradiated hemisphere (that facing the white-dwarf and boundary~layer)
of the mass donor. The second `S-wave' feature is travelling in phase
with the general trend of the H$\alpha$ line, thus following the
motion of the accretion disc and white dwarf. It appears in the
Doppler map as a smeared blob in the lower-central region. SS~Cyg has
a similar Doppler map (column~three of Fig.~\ref{fig:dopp}). There is a
prominent contribution from the hemisphere of the mass~donor facing
the white dwarf. In addition, there is a smeared blob here too,
appearing this time in the bottom-to-lower-right quadrant of the
Doppler map. The trailed spectra show a second `S-wave' component
moving in phase with the emission line, which apparently shifts from
negative to positive velocity around phase~0.4/0.5. Both of the
Doppler maps (for AH~Her and SS~Cyg) show emission at lower velocities
than the predicted outer disc velocities for an accretion~disc with a
radius $0.8R_{L_1}$ (plotted as dot-dashed circles on the Doppler
images), so-called ``sub-Keplerian'' velocities.

V426~Oph has a very peculiar H$\alpha$ map (see column~two of
Fig.~\ref{fig:dopp}). The emission appearing near zero velocity
doesn't all fit into the Roche lobe of the mass~donor. It appears to
be consistent with the velocity of the inner Lagrangian point, but
`spread out' in a semi-circular fashion.  The trail reconstructed from
the Doppler map appears to have more structure than can be clearly
seen in the actual data, however the feature seen in the actual data
at phase~0.4 isn't reproduced in the predicted data. There is also an
approximately steady emission varying at low velocities near the line
centre. This produces the semi-circular emission feature in the
Doppler image. The observed data also appear to show an eclipse at
phase~0. This has not been observed in V426~Oph before. Not easily
seen on the Doppler map is a region of emission at high velocities on
the left-hand side. This may be an indication of stream
overflow~\cite{stream}. The emission line behaviour is reminiscent of
the emission-line behaviour in the class of Nova-like variables,
sometimes called the SW~Sex stars~\cite{swsex}.

EM~Cyg (in column~one) again shows emission on the irradiated
hemisphere of the mass~donor (see North~et~al.~\shortcite{north} for
further discussion), and has the most ring-like feature indicative of an
accretion~disc. However, the peak velocities measured from this are
again apparently sub-Keplerian.

%%%%%%%%%%%%%%%%%%%%%%%%%%%%%%%%%%%%%%%%%%%%%%%%%%%%%%%%%%%%%%%%%
\section{Discussion}
\label{sec-dis}

Kolb \& Stehle~\shortcite{kolste} explain that the velocity dispersion
($\sigma$) is defined as the rms of the peculiar stellar
space velocity, $v$ in a local Galactic coordinate system (U,V,W),
where U points towards the Galactic centre, V is in the direction of
the Galactic rotation, and W is in the direction of the North Galactic
pole.  Wielen~et~al.~\shortcite{wie} (introduced by
Wielen~\shortcite{wie1}) deduced that the total dispersion,
$\sigma_{\rm v}$ is defined by
\begin{equation}
\sigma^{2}_{\rm v} = \sigma_{V}^{2} + \sigma_{U}^{2} + \sigma_{W}^{2}
\end{equation}
and obeys this empirical age--velocity relation:
\begin{equation}
\sigma_{\rm v} \simeq 10+21.5\biggl(\frac{t}{10^{9}{\rm
yr}}\biggr)^{1/2}\ \kms. 
\label{eq:disp}
\end{equation}
According to more recent studies there is some uncertainty surrounding
this relation. For a short discussion on this point see
Kolb~\shortcite{kolb}. In any case, this relation is also expected to
hold for binary systems, unless an event occurs during the evolution
in which a significant fraction of the total mass of the system is
violently, asymmetrically ejected. In the case of CVs, the progenitors
do eject a significant fraction of their mass during the
common-envelope period, but this occurs on a time scale much longer
than the duration of the orbital period, and so can be considered to
be symmetric about the rotation axis of the binary. This means that
the space velocity of the system is not affected. Kolb \& Stehle note
that repeated nova eruptions might alter the space velocity of the
system. This would in effect {\it increase} the velocity dispersion of
CVs. In this case, equation~\ref{eq:disp} would underestimate the
total dispersion.

The characteristic velocity observed for the CV population is the
velocity of the centre of mass of the binary system. Assuming CVs are
isotropically distributed about the Galaxy, the observed
$\gamma$ velocity dispersion is related to $\sigma_{\rm v}$ by
\begin{equation}
\sigma^{2}_{\rm |v|} = <\gamma^{2}> = \frac{1}{3}\sigma^{2}_{\rm v}
\label{eq:gam}
\end{equation} 
%In transforming the CV population from a distribution in age, $t$ and
%orbital period, $P_{\rm orb}$ to one over $\gamma$ velocity and
%orbital period, $P_{\rm orb}$, a distribution is obtained of the root
%mean square, $\sigma_{\gamma}$, of the range of $\gamma$ velocities
%at a particular orbital period.
It is seen that the CVs above the period gap (young) show a
significantly smaller dispersion in $\gamma$. What is notable about
the $\sigma_{\gamma}(P_{\rm orb})$ distribution is that it is
essentially {\it independent} of the selection effects which plague
the original distribution, $n(P_{\rm orb})$. The observations
described here are a first step towards a statistically meaningful
sample of $\gamma$ velocity measurements, for objects both above and,
eventually, below the period gap in the distribution of CVs.

From the study of van Paradijs, Augusteijn \& Stehle~\shortcite{vanP},
the 47 systems with orbital periods above the period gap have an
average $\gamma$ velocity $\overline{\gamma} = 4.6\pm33.1$\,\kms. They
admit that part of the dispersion is due to measurement errors. For
the 72 non-magnetic CVs, they determine an uncertainty in $\gamma$ of
$19.7$\,\kms\ calculated from the rms average of the standard
deviations of the distributions of single $\gamma$ measurements of the
individual sources. This is too large to measure a theoretical
dispersion of $15$\,\kms\ with any confidence. Assuming that this
measurement error and the intrinsic dispersion add in quadrature gives
their result, $\sigma_\gamma = 26.6$\,\kms\ for longer period DN
(almost twice the predicted value from Kolb \& Stehle,~1996).

\begin{figure}
\hspace*{\fill}
\psfig{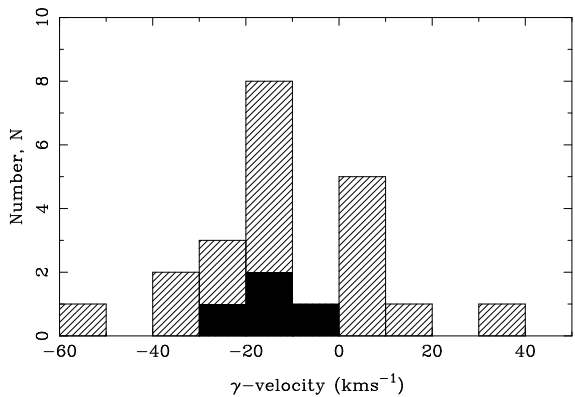}
\hspace*{\fill}
\caption{A histogram showing the spread of values for the
\protect{$\gamma$} velocities of the DN observed. The
solid histogram shows the results we have so far, and the area with hatched
lines denotes values for each of the observed systems gathered from
the literature.}
\label{fig:gamhist}
\end{figure}

Figure~\ref{fig:gamhist} compares the distribution of
$\gamma$ velocities as obtained a) in this paper (solid marking) and
b) from the literature (hatched lines) for the four DN observed in
this pilot study. We can see that the distribution is already
appearing much tighter, and the variation in the values obtained from
the literature is quite large. %It is obvious why, therefore, Kolb \&
%Stehle suggested obtaining a complete set of velocities for CVs both
%above and below the period gap. 
%In doing this we are directly testing
%the disrupted magnetic braking theory, a basic assumption in the
%standard model of CV evolution.

%In order to detect a velocity dispersion with a magnitude of $\approx
%15$\kms it is important to aim to obtain measurements of $\gamma$ to
%within $\approx 5$\kms. Using the methods detailed here, and knowing
%precisely the data-acquisition requirements needed during the
%observing run itself, we have formulated a process which specifically
%targets accurate measurements of absolute $\gamma$--velocities.  The
%requirement for optimum sampling of the spectra, in order to measure
%accurately the $\gamma$--velocities, reduces the potential use of
%archived data to carry out statistical analyses. In addition, adequate
%radial-velocity standards may not have been observed. The methods
%presented in this paper for measuring $\gamma$--velocities have been
%thoroughly tested here on four longer-period DN, and the initial
%results indicate that they are an accurate and simple way to determine
%values for CVs, minimising the potential sources of systematic error
%which exist.  
Initial results from the first sample of systems indicate that the
behaviour of the longer-period DN does follow closely that predicted
from the theory (as presented by Kolb \& Stehle, 1996). The initial
velocity dispersion value for the sample appears to be $\sim
8\kms$. We calculated a 95 per cent confidence limit on this result,
using the assumption that for a random sample ($X, S^2$) from a
population ($\mu, \sigma^2$), the quantity $NS^2/\sigma$ can be
described by a $\chi^2$ distribution, with $N-1$ degrees of freedom,
where $N$ is the number of measurements in the sample. For the 95 per
cent confidence interval
\begin{equation}
P\left(\chi^2_{0.025} < \frac{NS^2}{\sigma^2} < \chi^2_{0.975}\right)
= 0.95
\label{eq-prob}
\end{equation}
needed to be solved. Inverting the inequalities gives
%\begin{equation}
%\frac{NS^2}{\chi^2_{N-1}(0.975)} < \sigma^2_\gamma <
%\frac{NS^2}{\chi^2_{N-1}(0.025)} 
%\label{eq-95}
%\end{equation} which resulted in
the 95 per cent confidence interval $4 < \sigma_\gamma <
28$\,\kms. This range includes Kolb \& Stehle's theoretical
expectation ($15$\,\kms), but it also agrees with
van~Paradijs~et~al.'s result. We now need to increase the sample size
so as to conclusively reject or confirm the van Paradijs result. %If this is
%the case, it will highlight the problems with using $\gamma$ velocity
%values from the literature.  
As it stands, our result provides support for the theory of Kolb \&
Stehle~\shortcite{kolste}.%, now we must extend the analysis onto a
%statistically meaningful number of systems.

%%%%%%%%%%%%%%%%%%%%%%%%%%%%%%%%%%%%%%%%%%%%%%%%%%%%%%%%%%%%%%%%%%%%
\section{Conclusions}

Four long period ($P_{\rm orb} > 5$ hours) dwarf~novae have been
observed over a complete orbital period, in order to determine
accurate absolute systemic velocities (to within $\pm5$\,\kms). On
this initial sample, the aim was to test thoroughly the methods being
used to obtain the velocities, to make sure that any possible
sources of error were effectively minimised. Sources of systematic
error are fairly numerous using these methods, however they are very
easy to identify and minimise, and in certain cases to eliminate
altogether. A large sample of radial-velocity standard stars were
observed, to ensure that so-called `template mismatch' did not
introduce any significant error. This enabled us to compile a list of
reliable standard stars to use for future observing runs for this
project. In addition, because of the high accuracy requirement for
measurement of the systemic velocities, accurate measurements of other
binary parameters can be made. For example, high quality
radial-velocity curves mean that good semi-amplitude measurements can
be obtained. Existing orbital periods can also be checked, and a value
for the projected rotational velocity of the mass~donor can be
determined, implying that a value for the mass ratio, $q$, can be
calculated. This value of $q$ is then independent of the
semi-amplitude value determined from the emission lines (measurements
of which are uncertain and prone to error), and can then be used to
determine the component masses. Doppler maps have been constructed
from the H$\alpha$ line profiles obtained in this study. Due to the
coverage of a full orbital period, an orbit-averaged Doppler image has
be created for each object. These have proved surprising.  Further
evidence for the existence of ``sub-Keplerian'' velocities is seen in
the maps, as are peculiar features which may be be attributable to an
outflow from the systems. These features may be peculiar to H$\alpha$
emission; further investigation is needed.

Our initial results suggest that the observations agree with the
theory (as set out by Kolb \& Stehle, 1996). The observed velocity
dispersion for the sample is $\sim 8 \kms$, which is smaller than the
predicted value of $15 \kms$. If, with further observations, this
value does not alter significantly, then we can infer that the longer
period systems are a young subgroup, with an average predicted age
of $\le 1.5 {\rm Gyrs}$.
%Obviously, further measurements
%of the systemic velocities of more longer period DN are
%necessary. 
Once systemic velocities for the longer-period systems have been
adequately constrained then our understanding of their kinematics and
evolution will be more complete, and we can begin the search for the
predicted significant difference between the velocity dispersions of
CVs below and above the period gap.

\section*{Acknowledgments}
The authors wish to thank the referee, Hans Ritter, for drawing our
attention to  the SS~Cyg parallax.
RCN acknowledges a studentship from the University of Southampton and
a PPARC PDRA during the course of this work. The data reduction and
analysis were carried out on the Southampton node of the UK STARLINK
computer network. The INT is operated on the island of La Palma by the
Isaac Newton Group in the Spanish Observatorio del Roque de los
Muchachos of the Instituto de Astrofisica de Canarias.

\end{document}